\RequirePackage{lineno} 
\documentclass[useAMS,usenatbib]{mn2e}
\usepackage{graphicx}
\usepackage{longtable}
\usepackage{multirow}
\usepackage{txfonts}
\usepackage{color}
\bibpunct{(}{)}{;}{a}{}{,}

\def\gthree{G330.2$+$1.0}
\def\gone{G1.9$+$0.3}

\def\gammaray{$\gamma$-ray}
\def\gammarays{$\gamma$-rays}
\def\hess{H.E.S.S.}
\def\chandra{\emph{Chandra}}
\def\xmm{\emph{XMM-Newton}}
\def\asca{\emph{ASCA}}

\renewcommand{\deg}{\mbox{\ensuremath{^\circ}}} 
\def\hone{{H}{I}}

\title[TeV \gammaray\ observations of SNRs \gone\ and \gthree]{TeV $\gamma$-ray observations of the young synchrotron-dominated SNRs \gone\ and \gthree\ with \hess}

\author[H.E.S.S. Collaboration]{
\parbox{\textwidth}{\footnotesize H.E.S.S. Collaboration,
 A.~Abramowski,$^{1}$
 F.~Aharonian,$^{2,3,4}$
 F.~Ait Benkhali,$^{2}$
 A.G.~Akhperjanian,$^{5,4}$
 E.~Ang\"uner,$^{6}$
 G.~Anton,$^{7}$
 S.~Balenderan,$^{8}$
 A.~Balzer,$^{9,10}$
 A.~Barnacka,$^{11}$
% ...
}\vspace{0.6cm}\\
\parbox{\textwidth}{\scriptsize
$^1$ Universit\"at Hamburg, Institut f\"ur Experimentalphysik, Luruper Chaussee 149, D 22761 Hamburg, Germany \\
$^2$ Max-Planck-Institut f\"ur Kernphysik, P.O. Box 103980, D 69029 Heidelberg, Germany  \\
$^3$ Dublin Institute for Advanced Studies, 31 Fitzwilliam Place, Dublin 2, Ireland  \\
$^4$ National Academy of Sciences of the Republic of Armenia, Yerevan   \\
$^5$ Yerevan Physics Institute, 2 Alikhanian Brothers St., 375036 Yerevan, Armenia  \\
$^6$ Institut f\"ur Physik, Humboldt-Universit\"at zu Berlin, Newtonstr. 15, D 12489 Berlin, Germany  \\
$^7$ Universit\"at Erlangen-N\"urnberg, Physikalisches Institut, Erwin-Rommel-Str. 1, D 91058 Erlangen, Germany  \\
$^8$ University of Durham, Department of Physics, South Road, Durham DH1 3LE, U.K.  \\
$^9$ DESY, D-15736 Zeuthen, Germany  \\
$^{10}$ Institut f\"ur Physik und Astronomie, Universit\"at Potsdam,  Karl-Liebknecht-Strasse 24/25, D 14476 Potsdam, Germany  \\
$^{11}$ Nicolaus Copernicus Astronomical Center, ul. Bartycka 18, 00-716 Warsaw, Poland  \\
% ...
}}
\begin{document}

\date{Accepted 2014 March 6. Received 2014 March 6; in original form 2013 November 26.}
\vspace{2 cm}

\pagerange{\pageref{firstpage}--\pageref{lastpage}} \pubyear{2014}

\label{firstpage}

\maketitle

\begin{abstract}
The non-thermal nature of the X-ray emission from the shell-type supernova remnants (SNRs) \gone\ and 
\gthree\ is an indication of intense particle acceleration in the shock fronts of both 
objects. This suggests that the SNRs are prime candidates for very-high-energy 
(VHE; $E > 0.1$\,TeV) \gammaray\ observations. \gone, recently established as the youngest known
SNR in the Galaxy, also offers a unique opportunity to study the earliest stages of SNR evolution 
in the VHE domain. The purpose of this work is to probe the level of VHE \gammaray\ emission 
from both SNRs and use this to constrain their physical properties. Observations were conducted with the 
\hess\ (High Energy Stereoscopic System) Cherenkov telescope array over a more than six-year 
period spanning 2004--2010. The obtained data have effective livetimes of 67\,h for \gone\ and 16\,h
for \gthree. The data are analyzed in the context of the multi-wavelength observations currently 
available and in the framework of both leptonic and hadronic particle acceleration scenarios. No significant 
\gammaray\ signal from \gone\ or \gthree\ was detected.  Upper limits (99\% confidence level) to the TeV flux 
from \gone\ and \gthree\ for the assumed spectral index $\Gamma = 2.5$ were set at 
$5.6\times10^{-13}$\,cm$^{-2}$\,s$^{-1}$ above 0.26\,TeV and 
$3.2\times10^{-12}$\,cm$^{-2}$\,s$^{-1}$ above 0.38\,TeV, respectively. 
In a one-zone leptonic scenario, these upper limits imply lower limits on the interior 
magnetic field to $B_{\rm{G1.9}} \gtrsim 11$\,$\mu$G for \gone\ and to $B_{\rm{G330}} \gtrsim 8$\,$\mu$G 
for \gthree. 
In a hadronic scenario, the low ambient densities and the large distances to the SNRs
result in very low predicted fluxes, for which the \hess\ upper limits are not constraining.
\end{abstract}
 

\section{Introduction}
Supernova remnants (SNRs) are believed to be sites of efficient 
particle acceleration and are
expected to produce very-high-energy 
(VHE; $E > 0.1$\,TeV) \gammarays\ through the interaction of accelerated, high-energy particles with 
ambient medium and fields. TeV \gammaray\ emission is currently detected from a number of SNRs.
Of particular interest are those 
SNRs whose X-ray spectra are dominated by non-thermal emission 
such as 
RX\,J1713$-$3946 \citep{J1713Nature,Aharonian06RXJ1713,Aharonian07RXJ1713},
RX\,J0852.0$-$4622 (Vela Jr.) \citep{Aharonian05VelaJr,Aharonian07VelaJr}, and SN\,1006 \citep{sn1006}.
Synchrotron emission from these SNRs reveals the existence of high-energy 
electrons which implies that intensive particle acceleration is occurring at 
their shock fronts. It makes these sources particularly interesting for \gammaray\ astronomy
since high-energy particles accelerated at shock fronts can produce 
VHE \gammarays\ through the inverse Compton (IC) scattering of relativistic electrons on 
ambient photon fields, through Bremsstrahlung radiation of relativistic electrons, 
and through proton-nucleus interactions, and subsequent $\pi^0$ decay.

In this paper, the results of \hess\ observations of two other SNRs with dominant non-thermal 
X-ray emission, \gone\ \citep{reynolds08} and \gthree\ \citep{torii06}, are 
presented. 

The paper is organized as follows: In \textsection 2,
the general properties of \gone\ and \gthree, 
based on radio and X-ray observations, are presented. 
The \hess\ data analyses and results are described in \textsection 3. 
In \textsection 4, the non-detection of the SNRs is discussed in the context 
of leptonic and hadronic particle acceleration scenarios.
Finally, the conclusions are summarized in \textsection 5.

\begin{table*}
\centering
\caption{\hess\ observations of SNRs \gone\ and \gthree.}
\label{data}
\begin{tabular}{c c c c c c}
\hline
\hline
\\
SNR & Observation period & Livetime & Median offset angle & Median zenith angle & Threshold energy \\
\hline
\\
\gone\   & March 2004 -- July 2010 & 67\,h        & $1.3\deg$ & $16\deg$ & 0.26\,TeV \\
\gthree\ & June 2005 -- May 2009          & 16\,h & $1.6\deg$ & $30\deg$ & 0.38\,TeV \\
\hline
\end{tabular}
\end{table*}

\section{The young SNRs \gone\ and \gthree}

\subsection{\gone}

In 1984, a radio survey using the Very Large Array (VLA) at 4.9\,GHz
led to the discovery of \gone\ (also G1.87$+$0.33),
identified as an SNR based on its shell-like morphology and non-thermal radio emission \citep{green84}.  
\gone\ had the smallest angular extent ever measured for a Galactic SNR ($\sim$1.2\arcmin)
suggesting a young age $\lesssim 10^3$\,y and/or a large distance.
Further evidence for the youth of \gone\ came from VLA observations at 1.5\,GHz from 1985 \citep{green04} which 
clearly showed a circular symmetry, as observed in other young SNRs.

More recent observations at both X-ray \citep{reynolds08} and radio \citep{green08} wavelengths confirmed the
young age of \gone\ by directly measuring the expansion of the SNR since earlier epochs.
A spectral analysis of the \chandra\ X-ray data \citep{reynolds08, reynolds09} revealed that the spatially
integrated X-ray emission between 1.5 and 6 keV is well described as
synchrotron emission from an electron distribution characterized by a power-law with an exponential cut-off.
In the context of the \texttt{srcut} model
\footnote{The \texttt{srcut} model adopted by \citet{reynolds09}
describes the synchrotron radiation from an electron
distribution described by a power law with an exponential cut-off in a uniform magnetic field.} taking into account 
the effects of dust scattering,
a roll-off frequency
$\nu_{\rm{roll}} = 5.4^{+4.8}_{-2.4} \times 10^{17}$\,Hz (errors represent 90\% confidence limits),
one of the highest values ever reported for an SNR, and a spectral index
$\alpha = 0.634^{+0.021}_{-0.020}$ (90\% confidence limits; flux density $S$ scales with frequency as S$_{\nu}$ $\propto$ $\nu^{-\alpha}$)
were obtained, as well as the absorption column density
$N_{\rm{H}} = 3.48^{+0.87}_{-0.80} \times 10^{22}$\,cm$^{-2}$ \citep{reynolds09}. This fit was performed assuming a 1\,GHz flux density of 1.17\,Jy which is obtained by extrapolating the value at 1.5\,GHz for the observed $\alpha = 0.62$ \citep{reynolds09}. The estimate of the column density, together with the angular proximity of \gone\ to 
the Galactic Center, suggests a distance of $\sim$8.5\,kpc, which is
assumed throughout this paper.

The \chandra\ image further revealed that the shell had significantly expanded (by $\sim$16\%) to its present diameter
of 1.7\arcmin\ \citep{reynolds08}.  
An age $\lesssim$150\,y was then derived by comparing radio observations from 1985 and \chandra\ observations from 2007
\citep{reynolds08} and later confirmed using only radio observations from the VLA at two different epochs
\citep{green08,murphy08}.
These observations also imply a mean physical radius of $\sim$2\,pc and a mean expansion velocity of 
$\gtrsim$12\,000\,km\,s$^{-1}$ at the assumed distance of 8.5\,kpc \citep{green08}.
The most recent X-ray measurements by \cite{Carlton11} are in agreement,
finding an age $(156 \pm 11)$\,y assuming no deceleration has taken place, 
with a true age most likely being $\sim$110\,y.

The combined radio / X-ray image \citep{reynolds08}
shows a bright, nearly circular ring with extensions (``ears'') extruding symmetrically
from the East and West.
However, the radio and X-ray morphologies differ significantly from each other; 
while the radio source exhibits its maximum brightness
in the North, the X-ray source has a marked bilateral E-W symmetry
which includes the aforementioned X-ray "ears" not seen in at radio wavelengths.
Interaction of the SNR shock front with a roughly uniform magnetic field B could explain the 
bilateral X-ray morphology, 
provided that the electron acceleration is dependent on the obliquity angle between the shock normal and 
B \citep{reynolds09,Fulbright90},
but suggests that the large-scale B may not be important for the radio emission \citep{green08},
which exhibits a markedly different morphology. An alternative explanation for the bilateral X-ray 
morphology is that the proton injection rate is dependent on the obliquity angle.  This would result in magnetic field 
amplification being confined to the polar regions and is considered plausible for the related case of  
SNR SN\,1006 which also features bilateral morphology \citep[see e.g.][]{voelk2003}.
Recently, thermal X-ray emission was also discovered from the interior of the remnant and rim \citep{borkowski10}.
The featureless, non-thermal, synchrotron-dominated, X-ray spectrum of the integrated emission \citep{reynolds08,reynolds09}
implies electrons are efficiently accelerated, reaching a maximum (cut-off) energy $E_{\mathrm{cut}} = 58 (B / 10$\,$\mu$G)$^{1/2}$\,TeV.

For a sphere of radius 2.2\,pc, a Type Ia SN explosion model with an exponential ejecta profile 
\citep{dwarkadas&chevalier}
predicts an age of 100\,y and an ISM number density of about 0.04\,cm$^{-3}$ \citep{reynolds08}.
\citet{ksenofontov10} derive slightly different values of the age (80\,y) and number density ($\sim$0.02\,cm$^{-3}$),
assuming an expansion velocity of 14\,000\,km\,s$^{-1}$ and radius of 2\,pc in their 
diffusive shock acceleration (DSA) model. Studying the expansion of \gone\ by comparing \chandra\ 
X-ray images taken in 2007 and 2009, \citet{Carlton11} derived an ISM density of 0.022\,cm$^{-3}$ in agreement with \citet{ksenofontov10}.

\subsection{\gthree}

The radio source \gthree\ was identified as a Galactic SNR \citep{clark73,clark75}
on the basis of its non-thermal spectrum and its proximity to 
the Galactic plane.  Following observations at radio frequencies \citep{caswell83} 
showed the clumpy, possibly distorted, shell-like structure of the remnant delineated by eight "blobs" of elevated
brightness. They also showed the existence of a gradient in the surface brightness, with intensity higher towards the plane.
\citet{whiteoak&green96} classified 
\gthree\ as a possible composite-type SNR. The size of the shell is $\sim$11\arcmin\ 
in diameter \citep{caswell83, whiteoak&green96}.  

Based on \asca\ observations \citep{tanaka94}, \citet{torii06} discovered a featureless
X-ray spectrum between 0.7 and 10 keV with a photon index $\Gamma = 2.82^{+0.22}_{-0.21}$ and interstellar absorption 
$N_{\rm{H}} = 2.58^{+0.36}_{-0.34} \times 10^{22}$\,cm$^{-2}$. It was also fit with a power law with 
exponential cut-off (\texttt{srcut} model), deriving 
$\nu_{\rm{roll}} = 4.3 \times 10^{15}$\,Hz and $N_{\rm{H}} = 5.1 \times 10^{22}$\,cm$^{-2}$ 
\citep{torii06} for the fixed observed radio spectral index $\alpha = 0.3$ and 
flux density at 1 GHz of 5 Jy deduced from the source spectrum \citep{green04}. A general anti-correlation 
between radio and X-ray intensities was shown, explained by 
the different density of the interstellar medium (ISM) on the eastern and western 
sides of the remnant. 
Since the eastern shock is decelerating as it interacts with a denser ISM, 
electrons are accelerated to lower energies (GeV) than in the western shock.
Conversely, the western shock is interacting with an ISM of lower density,
resulting in acceleration to higher energies (TeV).
As a result, the X-ray emission is stronger 
in the western part of the shell and radio emission in the eastern part \citep{torii06}.
The lower limit on the 
distance $d_{\rm{G330}} \geq 4.9$\,kpc was calculated by \citet{mcclure01} using \hone\
absorption measurement. The distance to \gthree\ is assumed to be 5\,kpc hereafter.

Subsequent \chandra\ and \xmm\ observations \citep{park06, park09} revealed that 
the X-ray emission from \gthree\ is dominated by a power-law continuum ($\Gamma \sim 2.1$--2.5)
and comes primarily from thin filaments along the boundary of the shell.  
Measurements of the filament widths using \chandra\ images allow
the downstream magnetic field and maximum (cut-off) electron energy to be estimated as
$B \sim 14$--$20$\,$\mu$G and $E_{\mathrm{cut}} \sim 22$--$38$\,TeV,
respectively \citep{park09}.
\citet{park06} also discovered a point-like source, CXOU\,J160103.1$-$513353, at the center of the SNR,
claiming it to be a candidate central compact object (CCO). Additionally, 
evidence of pulsations was found with a period of $\sim$7.5\,s, although later \xmm\ observations 
\citep{park09} did not confirm this.  \chandra\ and \xmm\ observations also revealed  
faint, thermal X-ray emission in the eastern region of the shell of \gthree\ \citep{park09}.  
Using the thermal emission, the ISM 
density was calculated and appears to be low ($\sim 0.1$\,cm$^{-3}$). 
Assumptions on the ISM density and the distance to the SNR presented above lead to the 
estimation of the age of the remnant $t_{\rm{G330}} \simeq 1000$\,y according to the 
\citet{sedov} solution for the adiabatic stage of the hydrodynamical expansion of the SNR \citep{park09}.

\begin{table*}
\centering
\caption{Upper limits on the TeV \gammaray\ flux from SNRs \gone\ and \gthree.}
\label{UL}
\begin{tabular}{c c c c c c c c}
\hline
\hline
\\
         & N$_{\mathrm{ON}}$ & N$_{\mathrm{OFF}}$ & $\alpha$    & Excess       & Significance        & $F$ [cm$^{-2}$\,s$^{-1}$] \\
\hline
\\
         &                &                 &              &              &                      & $F (> 0.26\,\mathrm{TeV}) < 4.9 \times 10^{-13}$ for $\Gamma = 2.0$ \\
\gone    & 785     & 20537    & 0.038 & 6.4   & 0.2\,$\sigma$ & $F (> 0.26\,\mathrm{TeV}) < 5.6 \times 10^{-13}$ for $\Gamma = 2.5$ \\
         &                &                 &              &              &                      & $F (> 0.26\,\mathrm{TeV}) < 6.4 \times 10^{-13}$ for $\Gamma = 3.0$ \\
\hline
\\
         &                &                 &              &              &                      & $F (> 0.38\,\mathrm{TeV}) < 2.5 \times 10^{-12}$ for $\Gamma = 2.0$ \\
\gthree  & 874     & 10445    & 0.074 & 100.5 & 3.4\,$\sigma$ & $F (> 0.38\,\mathrm{TeV}) < 3.2 \times 10^{-12}$ for $\Gamma = 2.5$ \\
         &                &                 &              &              &                      & $F (> 0.38\,\mathrm{TeV}) < 3.9 \times 10^{-12}$ for $\Gamma = 3.0$ \\
\hline
\end{tabular}
\end{table*}

\section{Observations and analysis}
\subsection{The \hess\ telescopes}

\hess\ (High Energy Stereoscopic System) is an array of four, 13-m diameter, imaging atmospheric Cherenkov telescopes
(IACTs)
located in the Khomas Highland of Namibia at an altitude of 1800\,m above sea level \citep{bernloehr03,funk04}.
The telescopes have a nominal field-of-view (FoV) of 5\deg\ and are optimized 
for detecting \gammarays in the range $\sim$0.1\,TeV to $\sim$30\,TeV.
The angular resolution of the system is $\lesssim$0.1\deg\ and the average energy resolution is $\sim$15\% \citep{crab}.
The \hess\ array is capable of detecting point sources with a flux of $\sim$1\% of the Crab Nebula flux at the significance
of 5\,$\sigma$ in $\sim$10\,h at low zenith angles \citep{Ohm09}. 

\subsection{Data and analysis techniques}

\gone\ is located $\sim$2\deg\ from the supermassive black hole Sgr\,A$^{*}$ at the Galactic Center (GC) 
and the TeV \gammaray\ source HESS\,J1745$-$290 which is coincident with the position of both Sgr\,A$^{*}$ and the
pulsar wind nebula G359.95$-$0.04 \citep{sgrA_HESS}. 
Analyses of the SNR therefore benefit from the deep \hess\ exposure in the region.
More than half of the observations used for the analysis are obtained from Sgr A$^{*}$ 
observations, while the remainder is from the \hess\ Galactic Plane Survey 
\citep{Aharonian06Survey,Carrigan13}.
In order to reduce the large exposure gradient
towards the GC, only those observations centered within 1.5\deg\ from the 
\gone\ center were selected for the analysis.
The observations which pass the standard \hess\ data quality selection \citep{crab}  
span a six-year period from 2004 until 2010, have a livetime of 67\,h, and a median offset of $1.3\deg$
from \gone\ (see Table~\ref{data}).
For optimal spectral reconstruction,
the strict selection excludes observations taken during poor or variable weather conditions and includes only those
where all four telescopes were in operation.
The median zenith angle (ZA) is relatively low, 
$16\deg$,
leading to a low-energy threshold of 0.20\,TeV for individual \gammarays.
The analysis is performed above the \emph{safe energy threshold} of the cumulative
\gammaray\ dataset
(here, 0.26\,TeV) to avoid known biases in the reconstructed energy close to the threshold \citep{crab}.

Since the SNR has a diameter of $\sim$1.7\arcmin\ when observed at both radio and X-ray energies, 
and since the \hess\ point spread function (PSF) (68\% containment) is much larger ($\sim$10\arcmin\ diameter), 
the test region from which the signal is measured (ON region) was defined \emph{a priori} as 
a circular region with a radius of 0.10\deg, the standard size used to search for point-like sources with H.E.S.S.
The test region is positioned at the center of \gone\ 
at $\alpha_{\mathrm{J2000}} = 17^{\mathrm{h}}48^{\mathrm{m}}44^{\mathrm{s}}$, $\delta_{\mathrm{J2000}} = -27\degr09\arcmin57\arcsec$ \citep{green84}.

There is no other source present within the same \hess\ FoV of \gthree\ and it has less exposure than \gone. 
All available data from 2005 through 2009 within 2.5\deg\ of the center of the remnant were used for the analysis. 
It results in $\sim$16\,h of livetime using only data which passed standard \hess\ quality selection and includes only those
observations where at least three telescopes were in operation.
The data were taken at a median ZA of $30\deg$; 
the higher ZA results in a respectively higher energy threshold, 0.38\,TeV, compared to \gone.
The median offset of the observations is $1.6\deg$.
The datasets used for the analyses of both \gone\ and \gthree\ are summarized in Table~\ref{data}.

The size of \gthree\ is similar to the \hess\ PSF. Thus, 
in order to take into account all the emission from the remnant a bigger 
ON region as compared to \gone\ was chosen \emph{a priori}, defined as 
a circle with radius $0.22\deg$.
The test region is positioned at the center of the SNR at 
$\alpha_{\mathrm{J2000}} = 16^{\mathrm{h}}01^{\mathrm{m}}3.14^{\mathrm{s}}$, $\delta_{\mathrm{J2000}} = -51\degr33\arcmin54\arcsec$.

The \hess\ standard analysis\footnote{H.E.S.S. Analysis Package (HAP) version 11-02-pl07} 
\citep{crab} was used for the processing of extensive air shower (EAS) data from both \gone\ and \gthree\ observations.
The \emph{boosted decision trees method} (BDT), a decision-tree-based machine-learning algorithm \citep{Ohm09},
was used for $\gamma$-hadron separation, i.e. to select \gammaray-like events while reducing the hadronic background component.
The recorded EAS images were required to have integrated intensities per image of at least 80\,photoelectrons 
(p.e.; \emph{standard} cuts) in order to be included in the analysis.  
The relatively low cuts used on the EAS image intensities (compared to \emph{hard} cuts at, e.g., 200\,p.e.) 
allowed the inclusion of fainter EASs to probe the low-energy end of the VHE \gammaray\ spectra from both \gone\ and \gthree.
Over the six-year observation period,  
the optical reflectivity of the \hess\ telescope mirrors varied and the gains of the cameras' photomultiplier tubes changed. 
This time-dependent optical response was taken into account in the spectral reconstructions by calibrating the energy of each event
with EAS images of single muon rings passing close to the telescopes \citep{Bolz04,HESSICRC07}.

The \emph{reflected region background method} \citep{berge07} was used for background subtraction when measuring the VHE \gammaray\ flux
from both SNRs.
In this method, both ON and background (OFF) regions are identical in size and have identical offsets from the camera center,
such that they are affected by the radially-varying acceptance in the same manner. 
Nearby regions with known VHE \gammaray\ emission, including the diffuse emission near the GC, 
were excluded from all OFF regions in order to avoid contaminating the background estimation.

Results were cross-checked using the alternative \emph{Model analysis technique}\footnote{ParisAnalysis software version 0-8-18} \citep{deNaurois09}
as well as an independent calibration of the raw data and quality selection criteria. The results obtained with these different analysis chains are consistent.

\subsection{Flux upper limits}

\begin{figure}
\centering
\resizebox{\hsize}{!}{\includegraphics{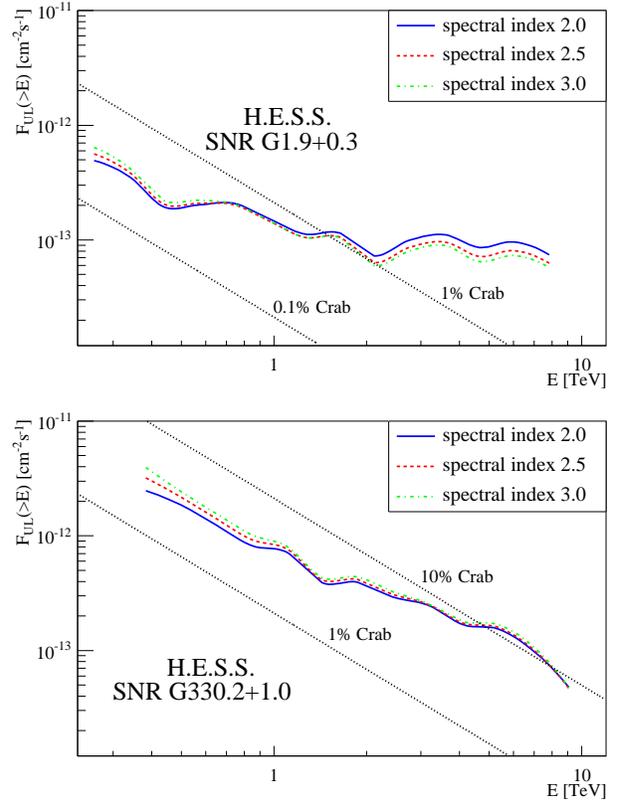}}
\caption{The upper limit (99\% confidence level) of the integrated TeV \gammaray\ flux from \gone\ (top) and \gthree\ (bottom)
for three different assumed spectral indices, $\Gamma = 2.0, 2.5$ and 3.0.}
\label{UL_G1.9_and_G330}
\end{figure}

Despite relatively deep exposures with the \hess\ telescopes, 
no significant VHE \gammaray\ signal was detected from \gone\ or \gthree.
The upper limits (ULs; 99\% confidence level) \citep{upper_limit} on the 
integral fluxes above the 0.26\,TeV (\gone) and 0.38\,TeV (\gthree) energy thresholds were calculated 
for three assumed spectral indices, $\Gamma = 2.0$, 2.5 and 3.0. 
The event statistics and ULs are summarized in Table \ref{UL}, 
where $N_{\mathrm{ON}}$ and $N_{\mathrm{OFF}}$ are numbers of 
ON and OFF region events, respectively, and $\alpha$ is the 
normalization factor between 
ON and OFF regions such that excess can be defined as  $N_{\mathrm{ON}} - \alpha N_{\mathrm{OFF}}$.  
The dependence of the integral flux UL on the energy threshold can be seen in Fig.~\ref{UL_G1.9_and_G330}.
Since the UL measurements are not strongly dependent on the value of $\Gamma$, 
ULs with assumed spectral index $\Gamma = 2.5$ are used hereafter in this paper.

\section{Discussion}

The synchrotron nature of the X-ray emission indicates that electrons in both SNRs are accelerated to very high (TeV) energies.
For such high energies, the acceleration process should run very similarly for electrons and hadrons. Some important differences
arise from the cut-off in the electron spectrum (due to electron radiation losses; see e.g. \citet{Reynolds_Keohane_99}) and in the number of
accelerated particles in each distribution. Nonetheless, the
existence of high-energy electrons directly shows that there should also exist hadrons accelerated to energies at least as high.

This leads to the expectation of \gammaray\ emission from inverse Compton (IC) scattering of relativistic electrons on photon fields
and/or from hadronic (e.g. proton-nucleus) interactions. 
The non-detection of this emission allows constraints to be placed on parameters
such as the magnetic field strength, the ISM density, the distance and the cosmic-ray 
(CR) efficiency, the latter defined as the fraction of SN explosion energy that is transferred to the particle acceleration.

\begin{table}
    \centering
    \caption{SED model fitting parameters.}
    \label{param}
    \begin{tabular}{ccccc}
      \\
      \hline\hline
      && \\
      SNR     & $\Gamma_{\mathrm{e}}$ & $B$             & $E_\mathrm{cut}$ & $W_\mathrm{tot}$    \\
              &          & [$\mu$G]        & [TeV]           & [erg]              \\
      \hline
      \multicolumn{5}{c}{Uncooled electron spectrum}\\
      \hline
      && \\
      \gone   & 2.2      & $>12.1$ & $<44$   & $<4.2\times10^{48}$ \\
      \gthree & 2.2      & $>8.0$  & $<21$   & $<13.2\times10^{48}$ \\ 
      \hline
      \multicolumn{5}{c}{Dominating synchrotron losses}\\
      \hline
      && \\
      \gone   & 2.0      & $>8.6$ & $<80$   & -- \\
      \gthree & 2.0      & $>4.3$ & $<56$   & -- \\ 
      \hline
    \end{tabular}
\end{table}

\subsection{Leptonic scenario}
\label{lept_scen}

Although the comparison of the X-ray and radio data reveals general anti-correlation for 
both SNRs indicating that radio and X-ray emitting electrons may not come from the same population, the one-zone 
leptonic model is used to obtain constraints on physical parameters of the remnants and ambient media.
Assuming that the radio and X-ray emission are produced by the same electron population via synchrotron radiation,
one can predict the $\gamma$-ray emission expected from the IC scattering of the same electrons on the cosmic 
microwave background (CMB) photons and other ambient photon fields. 
Although in the vicinity of the GC, the contribution of the infrared (IR) 
and optical photon fields to the resulting IC emission can be comparable to or even exceed the contribution 
from the CMB photons alone \citep{porter06}, it is very difficult to determine the interstellar 
radiation field at the location of a specific object.
Therefore, in this paper, we first consider 
CMB photons alone, since it is possible that there is no significant source of 
target photons in the proximity of \gone\ and \gthree, but then also discuss a 
potential contribution of the IR and optical photon fields to the overall IC emission 
and its impact on the resulting constraints on magnetic field and 
electron population parameters.

\begin{table}
\centering
\caption{Parameters of optical and IR photon fields.}
\label{photons}
\begin{tabular}{c c c c c}
\hline
\hline
&& \\
SNR&  \multicolumn{2}{c}{Optical photons}  &   \multicolumn{2}{c}{IR photons}\\
&&\\
        & $T_{\rm{opt}}$ & energy density & $T_{\rm{IR}}$ & energy density\\
        & [K]          &[eV\,cm$^{-3}$]  & [K]         & [eV\,cm$^{-3}$]\\
\hline
&&\\
\gone\   & 4300         & 14.6           & 48          & 1.5\\
\gthree\ & 3500         & 2.4            & 39          & 1.4\\
\hline
\end{tabular}
\end{table}

The spectral energy distribution (SED) for \gone\ and \gthree\ is calculated assuming the 
stationary case and the exponentially cut-off power-law distribution of the electron density with energies, 
\begin{equation}
{N_{\mathrm{e}}\left(\gamma\right) = K_{\mathrm{e}}\,\gamma^{-\Gamma_{\mathrm{e}}}\,e^{-\frac{\gamma}{\gamma_{\mathrm{cut}}}}, }
\end{equation}
where $\gamma$ is the electron Lorentz factor, $K_{\mathrm{e}}$ is the normalization, $\Gamma_{\mathrm{e}}$ is the spectral index, and $\gamma_{\mathrm{cut}} = E_{\mathrm{cut}} / \mathrm{m}_{\mathrm{e}}c^2$ is the cut-off Lorentz factor with the cut-off energy $E_\mathrm{cut}$ and the electron mass $\mathrm{m}_{\mathrm{e}}$. The synchrotron emission 
is calculated according to \citet{rybicki&lightman79} assuming the isotropic magnetic 
field and the isotropical distribution of the electron velocities. 
The correct integration over angle $\alpha$ between the electron velocity and the magnetic field is established 
using the function $G(x)$ introduced by \citet{AKP2010}. The IC emission is 
estimated according to \citet{blumenthal&gould} using the Klein-Nishina cross section.

\begin{figure}
\centering
\resizebox{\hsize}{!}{\includegraphics{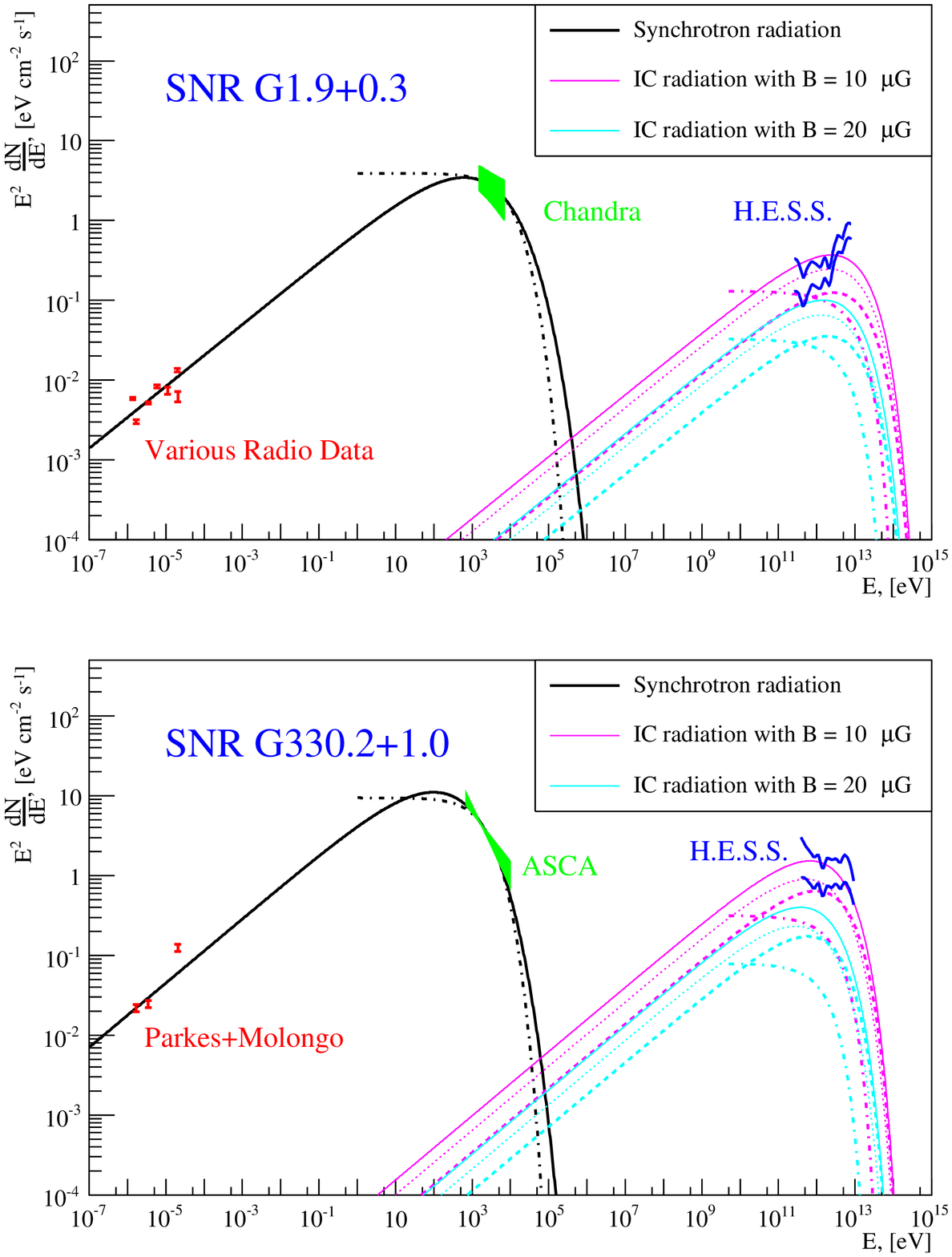}}
\caption{Spectral energy distributions of \gone\ (\emph{top}) and \gthree\ (\emph{bottom}) in a leptonic scenario.
The \hess\ upper limits on the differential flux are shown assuming two different spectral indices, 
2.0 (lower curve) and 3.0 (upper curve). 
The multi-frequency radio data shown for \gone\ was compiled by \citet{green08}; additional upper limits in the IR domain
\citep{Arendt89} are not shown because they lie outside of the plotted range and are not constraining. 
The solid and dot-dashed lines represent the modeled synchrotron and IC
emission spectra from uncooled and cooled (due to synchrotron losses) electron
spectrum, respectively. For the IC emission, dotted (resp. dashed) lines
correspond to the contribution due to IC scattering on CMB (resp. IR) photons,
in the case of the uncooled electron spectrum. The IC emission is calculated
for two assumptions on $B$. Note that the lower limit on the magnetic field is 
calculated comparing the integral upper limit on the \gammaray\ flux above the safe energy threshold to the model prediction 
of the flux above the same energy. See Section \ref{lept_scen} for details.}

\label{sed_G1.9_and_G330}
\end{figure}

In Fig.~\ref{sed_G1.9_and_G330}, SED models for \gone\ and \gthree\ are presented.
The IC contribution to the SED is presented for two different assumed values of the magnetic field $B$.
The synchrotron contribution to the SED (black solid lines) 
is modeled with the electron spectral 
index $\Gamma_{\mathrm{e}} = 2.2$ on both cases, which represents the multi-wavelength (MWL) 
observational data quite well. This electron spectral index corresponds to the radio spectral 
index of $\alpha = 0.6$. For \gthree, this value is very different from the observed spectral index of 0.3 reported 
by \citet{clark75} based on two observed points: at 408 MHz (Molongo Cross Telescope) and 5000 MHz (Parkes 64m radio telescope). However, 
subsequent observations at 843 MHz with the Molongo Observatory Synthesis Telescope \citep{whiteoak&green96} revealed a flux density which does not agree with such a low 
spectral index. The choice of $\alpha=0.6$ in this work is motivated by the necessity of fitting the X-ray data, which cannot 
be explained for $\alpha=0.3$ within this model.

Comparing the \hess\ integral flux ULs on the TeV $\gamma$-ray emission above the safe energy threshold (see Table~\ref{UL}; for assumed $\Gamma = 2.5$) to the predicted \gammaray\
flux above the same energy, within the context of the leptonic model presented above, one can calculate lower limits on the interior magnetic 
field strength $B$.  
The lower limits are found to be 12.1\,$\mu$G for \gone\ and 8.0\,$\mu$G for \gthree. 
Lower limits on $B$ in turn allow ULs on the electron cut-off energy, $E_\mathrm{cut}$, and the total energy in 
electrons, $W_\mathrm{tot}$, to be determined (see Table~\ref{param}).

Physical assumptions made in the model above are the same as in the \texttt{srcut} model for the synchrotron emission used 
to fit the X-ray data. Therefore, it might be useful to compare roll-off frequencies of the synchrotron spectrum of 
\gone\ and \gthree\ implied from this work with those obtained in the \texttt{srcut} fits in earlier studies. It should 
be noted though, that the srcut model is an approximation and is exact only for the radio spectral index $\alpha = 0.55$ 
(corresponding to the electron index $\Gamma_\mathrm{e} = 2.1$). The estimate of the $\nu_{\mathrm{roll}}$ can differ from the real value
by 20\% depending on the spectral index, and will be lower (resp. higher) for $\alpha<$ (resp. $>$) 0.55. 
The roll-off frequency $\nu_{\mathrm{roll}}$ is the the characteristic frequency of the photon emitted by the electron with 
the energy $E_\mathrm{cut}$ and it is given by \citep[with an error corrected]{Reynolds_Keohane_99}
\begin{equation}
  \nu_{\mathrm{roll}} = 1.6\times10^{16}\left(\frac{E_{\mathrm{cut}}}{10\hbox{ TeV}}\right)^{2}\left(\frac{B}{10\hbox{ $\mu$G}}\right) [\hbox{Hz}].
\end{equation}
For \gone, the roll-off frequency obtained in this work, $\nu_{\mathrm{roll,\,G1.9}} = 3.7\times10^{17}$\,Hz, is consistent with the one obtained 
in \citet{reynolds09}. In the case of \gthree, $\nu_{\mathrm{roll,\,G1.9}} = 5.6\times10^{16}$\,Hz is an order of magnitude higher than the one
derived by \citet{torii06}, which can be naturally explained by the different assumed spectral index: in \citet{torii06} the value of the 
radio spectral index was fixed to $\alpha = 0.3$, while in this work the synchrotron emission from \gthree\ is modeled for $\alpha=0.6$. 

The electron spectrum of the form of the power law with the exponential 
cut-off is valid only if the energy losses due to the synchrotron emission can 
be neglected. This regime is plausible for both \gthree\ and especially \gone\ 
due to their young age. The "break" energy above which synchrotron cooling 
starts to play an important role is given by the expression \citep{blumenthal&gould}

  \begin{equation}
    E_{\mathrm{syn}} = 1.3 \times 10^{3} \left(\frac{t_{\mathrm{age}}}{100\,\mathrm{y}}\right)^{-1} \left(\frac{B}{10\,\mu\mathrm{G}}\right)^{-2}\,\mathrm{TeV}.
  \end{equation} 

For the estimated ages of the SNRs and derived lower limits of the magnetic field 
upper limits on the break energy can be calculated resulting in $\sim900$ TeV for \gone\ and 
$\sim200$ TeV for \gthree. However, the higher magnetic field would significantly 
decrease the estimate of the break energy, i.e. a synchrotron cooling can occur. Significant 
synchrotron cooling modifies the shape of the initial electron spectrum obtained from the acceleration 
process. The modified electron spectrum is steepened by one and features a super-exponential cut-off 
\citep{ZA2007}:
  \begin{equation}
    N_{\mathrm{e}}(\gamma) \propto \gamma^{-(\Gamma_{\mathrm{e}}+1)} e^{-\left(\frac{\gamma}{\gamma_{\mathrm{cut}}}\right)^2}.
  \end{equation} 
Following a similar procedure as presented above for the case of the uncooled electron spectrum, 
the lower limit on the magnetic field and the upper limit on the cut-off energy can be estimated. The 
spectral index obtained in the particle acceleration is assumed to be $\Gamma_{\mathrm{e}} = 2$ and the radio data 
is not taken into account. In this scenario, the lower limits on magnetic field are $8.6\,\mu$G (29\% 
difference) for \gone\ and $4.3\,\mu$G (46\% difference) for \gthree. Upper limits on cut-off energies 
are 80 TeV (81\% difference) and 56 TeV (167\% difference) correspondingly.

To calculate the contribution of optical and IR photon fields (see Table~\ref{photons}),
the interstellar radiation field (ISRF) model of
\citet{porter06} was used. To simplify calculations ISRF models were
fit with Planck distributions for optical, IR and CMB photons.
For \gone, the adopted ISRF at $R = 0$\,kpc and $z = 0$\,kpc
was used, where $R$ is the distance from the GC and $z$ is the height above the Galactic plane.
For \gthree, the ISRF at $R = 4$\,kpc and $z = 0$\,kpc was adopted.
The ISRF at $R = 0$\,kpc and $z = 0$\,kpc can be described with an optical radiation at a
temperature $T_{\rm{opt}} = 4300$\,K with an energy density of $14.6$\,eV\,cm$^{-3}$ and 
a contribution from IR radiation at a temperature $T_{\rm{IR}} = 48$\,K with an energy density of
$1.5$\,eV\,cm$^{-3}$.
Similarly, the ISRF at $R = 4$\,kpc and $z = 0$\,kpc can be fit with the contribution from  
optical radiation at a temperature $T_{\rm{opt}} = 3500$\,K with an energy density of $2.4$\,eV\,cm$^{-3}$ and
a contribution from IR radiation at a temperature $T_{\rm{IR}} = 39$\,K with 
an energy density of $1.4$\,eV\,cm$^{-3}$. The contribution of the optical photons to the IC emission appears to be less than 1\% even in the 
relative vicinity of the GC and does not affect the derived constraints on the 
physical parameters presented in Table \ref{param}. In contrast, the inclusion of the 
IR photons in the modeling provide a significant effect on the results\footnote{An uncooled electron spectrum is assumed}. In this case the 
lower limits on the magnetic field are estimated to be 15.1\,$\mu$G (25\% difference) and 
10.5\,$\mu$G (31\% difference) for \gone\ and \gthree\ respectively. 
The higher the limits are on the magnetic field, the stronger the constraints are on
the cut-off energy and the total energy in electrons. For \gone, $E_\mathrm{cut}<40$\,TeV (10\% difference) and 
$W_\mathrm{tot}<3.0\times10^{48}$ erg (30 \% difference) and for \gthree, $E_\mathrm{cut}<18$\,TeV 
(14\% difference) and $W_\mathrm{tot}<8.5\times10^{48}$\,erg (36\% difference). In Fig. 
\ref{sed_G1.9_and_G330}, the contribution of the IR photons to the overall IC emission SED is 
shown with dashed lines.

The leptonic model of the broadband emission from \gone\ 
presented in this paper is similar to the purely leptonic model (in the 
test particle limit) considered by \citet{ksenofontov10}. The main difference 
is that \citet{ksenofontov10} assume a radio spectral index $\alpha = 0.5$, 
i.e. electron spectral index $\Gamma_{\mathrm{e}} = 2.0$, whereas in this paper the radio 
spectral index $\alpha = 0.6$ ($\Gamma_{\mathrm{e}}=2.2$) was adopted based on radio observations. Taking into 
account this difference, the results obtained by the two models are compatible. 
Nevertheless, given the low value obtained for the lower limit on $B$,
the purely leptonic scenario, with an unmodified shock and without magnetic 
field amplification, cannot be ruled out, in contrast to what was suggested by 
\citet{ksenofontov10}.

\subsection{Hadronic scenario}

The \hess\ ULs on the \gammaray\ flux from \gone\ and \gthree\ can also be 
compared to predictions based on a hadronic scenario,
where $\pi^0$ mesons would be created when CR ions accelerated in the 
supernova blast wave collide with the ambient thermal gas, producing \gammarays\
via $\pi^0$ decay. 
Since both SNRs exhibit synchrotron X-ray emission which reveals the 
existence of electrons with energies $\gtrsim 20$\,TeV, the maximum energy of accelerated 
hadrons should be at least 20 TeV. This suggests that the spectrum of $\gamma$-rays 
produced in proton-nucleus interactions extends up to at least a few TeV.
The expected VHE flux from an SNR in a hadronic scenario can be then described, according to \citet{drury94}, as

\begin{eqnarray}\nonumber
\label{hadronic}
F(> E) \approx && 8.84 \times 10^{6} q_{\gamma}(\geq \hbox{1\,TeV}) \left( \frac{E}{\hbox{1\,TeV}}\right)^{1-\Gamma_{\mathrm{p}}} \\
               && \theta \left(\frac{E_{\rm{SN}}}{10^{51}\hbox{\,erg}}\right)\left(\frac{d}{1 \hbox{\,kpc}}\right)^{-2}\left(\frac{n}{1\hbox{\,cm}^{-3}}\right) \hbox{\,cm}^{-2} \hbox{\,s}^{-1}
\end{eqnarray}

where $q_{\gamma}$ is the $\gamma$-ray emissivity normalized to the CR energy density,
$\Gamma_{\rm{p}}$ is the spectral index of the relativistic protons 
distribution, $\theta$ is the CR acceleration efficiency, and
$E_{\rm{SN}}$ is the SN explosion energy, $d$ is the distance to the SNR and $n$ is the 
ISM density. 
The emissivity $q_{\gamma}(\geq\hbox{1\,TeV})$ also depends on $\Gamma_{\rm{p}}$ (inversely proportional),
and \citet{drury94} have calculated $q_{\gamma}$ for spectral indices 2.1--2.7 \citep[see Table~1 in][]{drury94}, 
taking into account the contribution of nuclei other than H by multiplying the pure proton 
contribution by a factor of 1.5. The values $\Gamma_{\rm{p}} = 2.1$ and $q_{\gamma} = 1.02 \times 10^{-17}$ 
are adopted to predict the highest possible flux.
Furthermore, in this scenario, only emission from neutral pion decay is taken into account; charged pion
decay will contribute IC and Bremsstrahlung emission but with a much smaller contribution to the energetics.

After fixing the spectral index and the CR production rate, four parameters remain free: 
$\theta$, $E_{\rm{SN}}$, $d$ and $n$.  Assuming the explosion energy released is $10^{51}$\,erg and 
taking into account the estimated distance to the SNR, one can constrain the product of the CR efficiency and the ISM 
density using the \hess\ UL. The resulting \gammaray\ spectrum should roughly follow 
the energy spectrum of protons. Since $\Gamma_{\mathrm{p}} = 2.1$ is assumed, the \hess\ UL
with the assumed index of 2.0 should be used for placing constraints as the closest to the modeled 
\gammaray\ spectrum.

The expected flux above 0.26 TeV from \gone\ assuming $d = 8.5$\,kpc is then
\begin{equation}
\label{hadronicG1.9}
F_{\rm{G1.9}}(>\hbox{ 0.26 TeV}) \approx 5.5\times10^{-12} \theta_{\rm{G1.9}} \left(\frac{n_{\rm{G1.9}}}{1\hbox{ cm}^{-3}}\right) \hbox{ cm}^{-2} \hbox{ s}^{-1}.
\end{equation}  
The \hess\ UL on the flux above the same energy, $4.9 \times 10^{-13}\hbox{ cm}^{-2} \hbox{ s}^{-1}$, 
can be used to provide an UL on the product of the density and efficiency,
\begin{equation}
\theta_{\rm{G1.9}} \left(\frac{n_{\rm{G1.9}}}{1\hbox{ cm}^{-3}}\right) < 0.09.
\end{equation}

During the free expansion stage of the SNR's evolution, 
which \gone\ is assumed to be in, the 
CR efficiency $\theta$ is expected to be very low, $\theta \ll 1$ \citep{drury94}.
\citet{ksenofontov10} show that at the age of 100\,y, the CR efficiency for \gone\ should be about 
$3\times10^{-3}$. The typical value of the CR efficiency during the adiabatic stage of SNR 
evolution $\theta = 0.1$ can serve as ULs for the case of \gone. Here, the range of 
values $3\times10^{-3}\leq\theta_{\mathrm{G1.9}} \leq 0.1$ is considered.
This leads to an UL on the 
ISM density $n_{\rm{G1.9}} < (1 - 30)$\,cm$^{-3}$ depending on the assumed $\theta_\mathrm{G1.9}$.
This UL is 2--3 orders of magnitude higher than the estimate based on 
the Type Ia SN model of \citet{dwarkadas&chevalier}
and the \hess\ flux UL is therefore not constraining.
On the other hand, assuming the density $n_{\rm{G1.9}} \approx 0.04$\,cm$^{-3}$ \citep{reynolds08}, an UL on the CR efficiency can be obtained,
$\theta_{\rm{G1.9}} < 2.3$. Since $\theta$ is defined only in 
the range 0--1, this limit is also not constraining.

For SNR \gthree, the expected flux above 0.38\,TeV at the distance of 5\,kpc is  
\begin{equation}
\label{hadronicG330}
F_{\rm{G330}}(>\hbox{ 0.38 TeV}) \approx 10^{-11} \theta_{\rm{G330}} \left(\frac{n_{\rm{G330}}}{1\hbox{ cm}^{-3}}\right) \hbox{ cm}^{-2} \hbox{ s}^{-1}.
\end{equation} 
The \hess\ UL on the flux above this energy $2.5\times10^{-12}\hbox{ cm}^{-2}  \hbox{ s}^{-1}$ constrains 
the product of the CR efficiency and the density
\begin{equation}
\theta_{\rm{G330}} \left(\frac{n_{\rm{G330}}}{1\hbox{ cm}^{-3}}\right) < 0.25.
\end{equation}
It corresponds to an UL on the ISM density $n_{\rm{G330}}<2.5$ cm$^{-3}$, assuming 
the typical value of the CR efficiency during the adiabatic stage of SNR evolution,
$\theta_{\rm{G330}} = 0.1$, and to an UL on the CR 
efficiency $\theta_{\rm{G330}}<2.5$ assuming the \citet{park06} estimate on the 
ISM density $n_{\rm{G330}} \approx 0.1$\,cm$^{-3}$. 
In the case of \gthree, ULs estimated within the hadronic scenario are 
also not strongly constraining. Estimates of the ULs on the product of the 
CR efficiency and the density of both \gone\ and \gthree\ 
are within the range of estimates for a subset of 20 other SNRs recently studied 
by \citet{bochow}.

Alternatively, with existing estimates of the ISM densities and assumptions on CR efficiencies,
one can predict the expected fluxes from \gone\ and \gthree. For example, 
assuming $n_{\rm{G1.9}} = 0.04\,$cm$^{-3}$ and 
$\theta_{\rm{G1.9}} = (0.003 - 0.1)$, the expected VHE \gammaray\ flux from \gone\ above 0.26\,TeV according to Eq.~\ref{hadronicG1.9}
is in the range of $(0.07 - 2.2)\times10^{-14}\hbox{ cm}^{-2} \hbox{ s}^{-1}$, 1--3
orders of magnitude lower than the \hess\ UL.
For \gthree, assuming  $n_{\rm{G330}} = 0.1\,$cm$^{-3}$ and $\theta_{\rm{G330}} = 0.1$ according to Eq. \ref{hadronicG330} one 
can calculate the expected flux above 0.38 TeV of $1\times10^{-13}\hbox{ cm}^{-2} \hbox{ s}^{-1}$, 
25 times lower than the UL. 

Although the \hess\ ULs for both SNRs do not constrain the predictions of this scenario,
it should be noted that there exist non-negligible uncertainties in many of the model parameters.
In particular, the expected $\gamma$-ray flux is very sensitive to the estimate of 
the distance to the source. According to \citet{ksenofontov10}, the dependence 
of the $\gamma$-ray flux on the distance for \gone, taking into account the relations 
between the distance and the ISM density, SNR radius and shock velocity, is 
$F_{\gamma} \propto d^{-11}$. 
Therefore, even a small decrease in the distance estimate would significantly 
increase the expected flux and consequently improve the constraints 
on the ISM density and the CR efficiency. Specifically, a reduction of the distance to 
\gone\ by $46\%$ to 4.6\,kpc would increase the expected flux, calculated for the lowest assumed CR 
efficiency of $0.003$, to the level of the \hess\ UL.  
For \gthree, the expected flux scales simply as $d^{-2}$ and would be 
compatible with the \hess\ UL if the distance to the source were reduced by $25\%$, to 3.8\,kpc.

\section{Summary}

The SNRs \gone\ and \gthree\ can serve as valuable astrophysical laboratories for investigating the MWL properties of young,
shell-type SNRs whose emission is dominated by non-thermal synchrotron emission. Observations in different energy regimes can provide
insight on the physical properties of this important subclass of SNRs.
\hess\ observations 
in particular can provide a unique probe at the highest energies, in the TeV \gammaray\ 
regime.

Despite relatively deep exposures, the \hess\ data do not show any signs of significant TeV \gammaray\ emission from either SNR.
Consequently, the 99\% confidence level 
ULs on the TeV \gammaray\ flux from these sources were determined.
For assumed power-law spectra with a spectral index $\Gamma = 2.5$, the obtained ULs are 
$F_{\rm{G1.9}}(>0.26\hbox{ TeV}) < 5.6 \times 10^{-13}$ cm$^{-2}$s$^{-1}$ for \gone\ 
and $F_{\rm{G330}}(>0.38\hbox{ TeV}) < 3.2 \times 10^{-12}$ cm$^{-2}$s$^{-1}$ for \gthree. 

The ULs on the TeV \gammaray\ flux provide an opportunity to set constraints on the 
magnetic field in the context of a leptonic particle acceleration scenario and on the ISM density and 
CR efficiency in a hadronic scenario.  Lower limits on the interior magnetic fields 
were estimated at 12\,$\mu$G for \gone\ and 8\,$\mu$G for \gthree. 
The obtained lower limits can be satisfied without requiring magnetic-field amplification 
beyond simple compression. In the case of the 
hadronic scenario, the ULs are two orders of magnitude greater than the flux prediction.
Obtained ULs on the ISM densities are compatible with other estimates of the densities 
(from the thermal X-ray emission for \gthree\ and from the expansion rate for \gone).
The CR efficiency, however, cannot be significantly constrained with the current dataset.

The non-detection of \gone\ and \gthree\ in the TeV \gammaray\ domain 
can be understood by examining those characteristics which set them apart from 
other members of this subclass, notably
Vela Jr., RX\,J1713$-$3946, and SN 1006,
all of which have been previously detected by \hess\ to emit TeV \gammarays.
While most are situated at relatively near distances from the Sun ($d \lesssim 2$\,kpc), 
\gone\ and \gthree\ are both significantly farther away ($d \gtrsim 5$\,kpc).
Their remoteness considerably reduces the \gammaray\ flux, particularly in hadronic scenarios.
Higher ambient densities would also have increased the flux predictions in such a scenario.
Finally, the relatively young ages of these remnants are problematic due to 
smaller population of high-energy particles, which results in lower \gammaray\ flux.
In the leptonic scenario, this necessitates a low magnetic field to 
compensate and achieve a flux which is detectable with the current 
IACTs, and may even challenge next-generation instruments.
\gone\ is also unique due to its exceptionally young age in comparison to the other SNRs.  
This could imply that, at least for \gthree, the age is not the main problem and that it could have been detected if it were closer.

\gthree\ and \gone\ remain promising targets for \gammaray\ observations at TeV energies, in particular with the future
generation of instruments, namely the Cherenkov Telescope Array (CTA) due to its 
$\sim10$ times higher sensitivity \citep{cta}.

\section*{Acknowledgements}
{\footnotesize
We are very grateful to S. Reynolds for helpful discussions and for 
providing us with the power-law fit of the \gone\ X-ray data.

The support of the Namibian authorities and of the University of
Namibia in facilitating the construction and operation of H.E.S.S.\ is
gratefully acknowledged, as is the support by the German Ministry for
Education and Research (BMBF), the Max Planck Society, the French
Ministry for Research, the CNRS-IN2P3 and the Astroparticle
Interdisciplinary Programme of the CNRS, the U.K. Particle Physics and
Astronomy Research Council (PPARC), the IPNP of the Charles
University, the South African Department of Science and Technology and
National Research Foundation, and by the University of Namibia. We
appreciate the excellent work of the technical support staff in
Berlin, Durham, Hamburg, Heidelberg, Palaiseau, Paris, Saclay, and in
Namibia in the construction and operation of the equipment.
}

\bibliographystyle{mn2e}
\bibliography{G330_and_G1.9_mnras_rr2} 

\label{lastpage}
\end{document}